\documentclass[journal]{IEEEtran}
\usepackage{algorithm2e}
\usepackage{amsmath,graphicx}
\usepackage{amssymb}
\usepackage{array}
\usepackage{bbm}
\usepackage{booktabs}
\usepackage{cite}
\usepackage[utf8]{inputenc}
\usepackage{mathtools}
\usepackage{mystyle}
\usepackage{multirow}
\usepackage{tikz}
\usepackage{hyperref}
\usepackage{pgfplots}
\usepackage{stfloats}
\usepackage{pifont}

\usepackage{subcaption,caption}
\usepackage{textcomp}
\usepackage{tikz-3dplot}
\usepackage{listings,xcolor}
\usepackage{url}
\usepackage{verbatim}
\usetikzlibrary{arrows.meta}
\usetikzlibrary{positioning,through, backgrounds,decorations.pathreplacing,shapes.misc}
\pgfplotsset{compat=1.17}
\pgfplotsset{every axis/.append style={
                    label style={font=\Large},
                    tick label style={font=\Large}  
                    }}

\tdplotsetmaincoords{70}{30}

\definecolor{codegreen}{rgb}{0,0.6,0}
\definecolor{codegray}{rgb}{0.5,0.5,0.5}
\definecolor{codepurple}{rgb}{0.58,0,0.82}
\definecolor{backcolour}{rgb}{0.95,0.95,0.92}
\lstdefinestyle{lsty}{
    backgroundcolor=\color{backcolour},   
    commentstyle=\color{codegreen},
    keywordstyle=\color{magenta},
    numberstyle=\tiny\color{codegray},
    stringstyle=\color{codepurple},
    basicstyle=\ttfamily\footnotesize,
    breakatwhitespace=false,         
    breaklines=true,                 
    captionpos=b,                    
    keepspaces=true,                 
    numbers=left,                    
    numbersep=5pt,                  
    showspaces=false,                
    showstringspaces=false,
    showtabs=false,                  
    tabsize=1
}
\begin{document}

\title{Written Term Detection\\ Improves Spoken Term Detection}
\author{Bolaji Yusuf,~\IEEEmembership{Graduate~Student~Member,~IEEE},~and~Murat Saraçlar,~\IEEEmembership{Senior~Member,~IEEE}
\thanks{Bolaji Yusuf is with Boğaziçi University Department of Electrical and Electronics Engineering, 34342 Istanbul, Turkey and also with Brno University of Technology, Faculty of Information Technology, Speech@FIT, 612 00 Brno, Czechia (e-mail: bolaji.yusuf@boun.edu.tr)
}
\thanks{Murat Saraçlar is with Boğaziçi University Department of Electrical and Electronics Engineering, 34342 Istanbul, Turkey (e-mail:muratsaraclar@ieee.org)
}
}

\markboth{Journal of \LaTeX\ Class Files,~Vol.~14, No.~8, August~2021}%
{Shell \MakeLowercase{\textit{et al.}}: A Sample Article Using IEEEtran.cls for IEEE Journals}


\maketitle

\begin{abstract}
End-to-end (E2E) approaches to keyword search (KWS) are considerably simpler in terms of training and indexing complexity when compared to approaches which use the output of automatic speech recognition (ASR) systems.
This simplification however has drawbacks due to the loss of modularity.
In particular, where ASR-based KWS systems can benefit from external unpaired text via a language model, current formulations of E2E KWS systems have no such mechanism.
Therefore, in this paper, we propose a multitask training objective which allows unpaired text to be integrated into E2E KWS without complicating indexing and search.
In addition to training an E2E KWS model to retrieve text queries from spoken documents, we jointly train it to retrieve text queries from masked written documents.
We show empirically that this approach can effectively leverage unpaired text for KWS, with significant improvements in search performance across a wide variety of languages.
We conduct analysis which indicates that these improvements are achieved because the proposed method improves document representations for words in the unpaired text.
Finally, we show that the proposed method can be used for domain adaptation in settings where in-domain paired data is scarce or nonexistent.
\end{abstract}

\begin{IEEEkeywords}
keyword search, spoken term detection, keyword spotting, end-to-end keyword search, multitask learning, domain adaptation, masked language modeling.
\end{IEEEkeywords}

\section{Introduction}
\label{sec:intro}
Keyword search (KWS), known alternatively as spoken term detection, is a branch of spoken content retrieval task concerned with retrieving speech segments where a user-provided query is uttered.
Given a user's short written query, a KWS system searches an archive of speech and returns those utterances in the archive hypothesized to contain the query, timestamps showing the exact location of each hypothesis and a set of scores denoting the system's confidence in the hypotheses.

The traditional approach to KWS involves building a large vocabulary continuous speech recognition (LVCSR) system, using it to decode the archive, and, from the resulting lattices, constructing an inverted index in which queries are searched~\cite{saraclar2004lattice,Ng2000,szoke2008hybrid,can2011lattice,chelba2008retrieval}.
However, this approach inherits the shortcomings of the underlying automatic speech recognition (ASR), most notably, the non-trivial complexity and computational costs associated with ASR training and decoding.
There has therefore been interest in end-to-end (E2E) approaches which eschew the ASR part of the KWS pipeline.
These E2E systems are trained to directly predict whether, and where, a query occurs in a given speech segment, leading to a much simpler system in terms of training and search~\cite{audhkhasi2017end,yusuf21_interspeech,svec21_interspeech}.
Although E2E KWS systems, like the one in this paper, still rely on simple ASR systems to get timing information at training time, they feature a much more simplified indexing and search scheme, comparable in complexity to acoustic modeling in ASR.

Its complexities notwithstanding, ASR-based KWS still maintains some advantages over E2E KWS in terms of both efficiency and accuracy of search.
While ASR-based systems transcribe the archives into text-based structures such as factor transducers~\cite{can2011lattice}, confusion networks~\cite{mangu2014efficient} and position specific posterior lattices~\cite{CHELBA2007458} which allow fast, sub-linear indexes, E2E methods generally rely on inner-product search with fixed frame-rate vector document representations.
Thus, the storage and computational cost of E2E KWS grows linearly in the duration of the archive.
In addition to efficiency, ASR-based KWS also outperforms E2E KWS in terms of search accuracy.
The performance advantage of ASR-based KWS is especially pronounced for short queries while E2E KWS tend to have the advantage for longer queries~\cite{yusuf2023end}.
Nevertheless, the two approaches tend to be complimentary and prior work has achieved significant improvements in search accuracy by combining them across queries of all lengths~\cite{yusuf21_interspeech,svec21_interspeech,yusuf2023end}.

As with E2E systems in other domains, the simplification in E2E KWS comes at the expense of data efficiency as these systems generally require larger amounts of labeled training data than their more modular counterparts.
Of particular interest to us is that ASR-based systems (even end-to-end ASR systems) can be improved with unpaired text data independent of the paired training speech-text data.
This naturally raises the question of how to use large text-only corpora to improve E2E KWS systems.
Since E2E KWS systems, as have been explored in literature, model span probabilities and not word probabilities, they cannot make use of language models which constitute the primary method of using text-only data to improve ASR systems.
On the other hand, there has been a recent trend in E2E ASR of using joint training with text-to-text transduction tasks to integrate the unpaired text into ASR training and reduce the dependency on external language models during ASR inference~\cite{wang21t_interspeech,yusuf22usted,thomas2022integrating}.

Inspired by these approaches, in this paper, we propose training an E2E KWS system jointly with an auxiliary text-to-text task.
Taking the E2E KWS model of~\cite{yusuf2023end} as the baseline, we introduce a joint training scheme where, in addition to the baseline training of predicting the locations of short written queries in \textit{speech} segments, the model is also trained to predict the locations of written queries in masked \textit{written} sentences.
As this auxiliary training objective can be computed with purely textual inputs, it provides a way to incorporate text-only corpora into the KWS model.

We conduct extensive experiments which yield the following results:
\begin{itemize}
    \item The proposed model consistently and significantly improves keyword search performance across several languages, domains and input feature choices.
    Moreover, the proposed joint speech-text training scheme is orthogonal to multilingual pretraining and data augmentation, and can be used alongside them to achieve even better performance.
    \item The proposed joint training method improves document representation of phrases contained in the auxiliary unpaired text, both when such phrases exist in spoken form in the paired KWS training data and when they do not.
    \item Training with unpaired text from a domain improves performance on test sets in that domain, and therefore provides a viable solution for dealing with domain mismatches between the KWS train and test sets.
\end{itemize}

The rest of the paper is organized as follows:
Section~\ref{sec:related} covers previous related work;
Section~\ref{sec:methods} recapitulates the baseline end-to-end KWS framework which we build upon and then describes the proposed model;
Section~\ref{sec:experiments} details the experiments conducted and discusses the results of those experiments; Section~\ref{sec:conclusion} concludes the paper with a summary and future research directions.

\section{Related Work}
\label{sec:related}
Our work falls within the gamut of ASR-free KWS systems which attempt to simplify the KWS pipeline.
Some of the earlier approaches to this include the use of point-process models~\cite{jansen2009point,Kintzley} and dynamic time warping~\cite{gundougdu2017joint,gundogdu2018generative,hazen2009query}.
More recent approaches use neural architectures which encode queries and documents and effect search by combining those encodings~\cite{audhkhasi2017end,yusuf21_interspeech,fuchs2021cnn,svec21_interspeech}, with especially~\cite{yusuf21_interspeech} and~\cite{svec21_interspeech} achieving high efficiency by using completely separated encoders for the query and the document and combining the encodings by simple dot-products.
Several improvements have been made to the document representations including pretraining the document encoder as an autoencoder~\cite{audhkhasi2017end}, an ASR encoder~\cite{zhao2020end}, a self-supervised model~\cite{svec22_interspeech} and a multilingual KWS document encoder~\cite{yusuf2023end}.
However, none of these approaches have been able to integrate unpaired text directly into the keyword search model.
We note that~\cite{audhkhasi2017end} and~\cite{fuchs2021cnn} pretrain their \textit{query} encoders using external text.
Unlike those, we use the unpaired text to better train our \textit{document} encoder.
As we will show in Section~\ref{sec:experiments:layers}, using unpaired text to train the document encoder with our method leads to significantly better KWS performance than using it for training only the query encoder.

Classical ASR-based methods can easily incorporate unpaired text as they are modular.
Various works have shown that using external text can significantly improve ASR-based KWS by using such text to augment the ASR pronunciation lexicon~\cite{10.1145/1571941.1571958,chen2013quantifying} and the language model~\cite{gandhe2013using,mendels2015improving}.
These works show that the KWS improvements can be substantial even when improvements to the underlying ASR system are less pronounced, due to the effect on rare and out-of-vocabulary (OOV) queries.
The fundamental question of this paper is how to leverage such external text for end-to-end KWS methods which possess neither lexicons nor language models.

A related line of research is the use of unpaired speech for training.
This includes pretraining with surrogate unsupervised objectives on large, untranscribed corpora and then finetuning on paired data~\cite{Khurana2020Conv,baevski2020wav2vec,liu2021tera,babu22_interspeech},
or semi-supervised training which involves training a seed ASR model on small transcribed data, using it to transcribe otherwise unlabeled data and then adding the resulting automatically-transcribed data into the training pool for further training~\cite{vesely2013semi,kahn2020self,khurana2021unsupervised}.
The work most related to ours in this direction is~\cite{svec21_interspeech}, in which an ASR system was used to transcribe large quantities of speech for training E2E KWS.
Overall, these works, which improve performance by making use of unpaired speech, are orthogonal to ours which makes use of unpaired text.
In our experiments, we use the pretrained model from~\cite{babu22_interspeech} to extract input features and we show that adding unpaired text with our method yields consistent improvements.

A more related line of work involves using unpaired text data directly in ASR training.
One way of doing so is using a text-to-speech (TTS) system to generate matching speech, and including the resulting paired data as part of ASR training~\cite{rossenbach2020generating,wang2020improving,baskar2021eat}.
However TTS adds its own significant computational and modeling complexity.
Moreover, robust TTS systems are generally a luxury only available for high resource languages.
Therefore, joint speech-text models such as MMDA~\cite{renduchintala18_interspeech}, PSDA~\cite{wiesner19_interspeech}, MUTE-L~\cite{wang21t_interspeech}, USTED~\cite{yusuf22usted}, Textogram~\cite{thomas2022integrating} and MAESTRO~\cite{chen22r_interspeech} have gained interest as a way of integrating unpaired text into end-to-end ASR to improve performance on ASR, as well as other downstream tasks such as spoken language understanding~\cite{thomas2022towards} and spoken machine translation~\cite{tang-etal-2021-improving}.
These models incorporate unpaired text by treating the entire ASR model as part of a larger multimodal text generator, some of whose parameters can be jointly trained for text-to-text transduction without any explicit TTS synthesis.
By improving the underlying ASR system, these methods can plausibly be used to improve ASR-based KWS systems, especially recently proposed KWS systems based on end-to-end ASR~\cite{audhkasi2018E2eAsrKws,shi2021timestamp,yang2022keyword,10097249}.
However, they cannot work for end-to-end ASR-free KWS systems which are generally discriminators rather than text generators.
Our proposed method introduces a surrogate objective for incorporating text into the discriminative framework ASR-free KWS systems.

\section{Methods}
\label{sec:methods}
\begin{figure*}[t]
    \centering
    \begin{subfigure}[b]{0.27\textwidth}
        \centering
        \includegraphics[width=\linewidth]{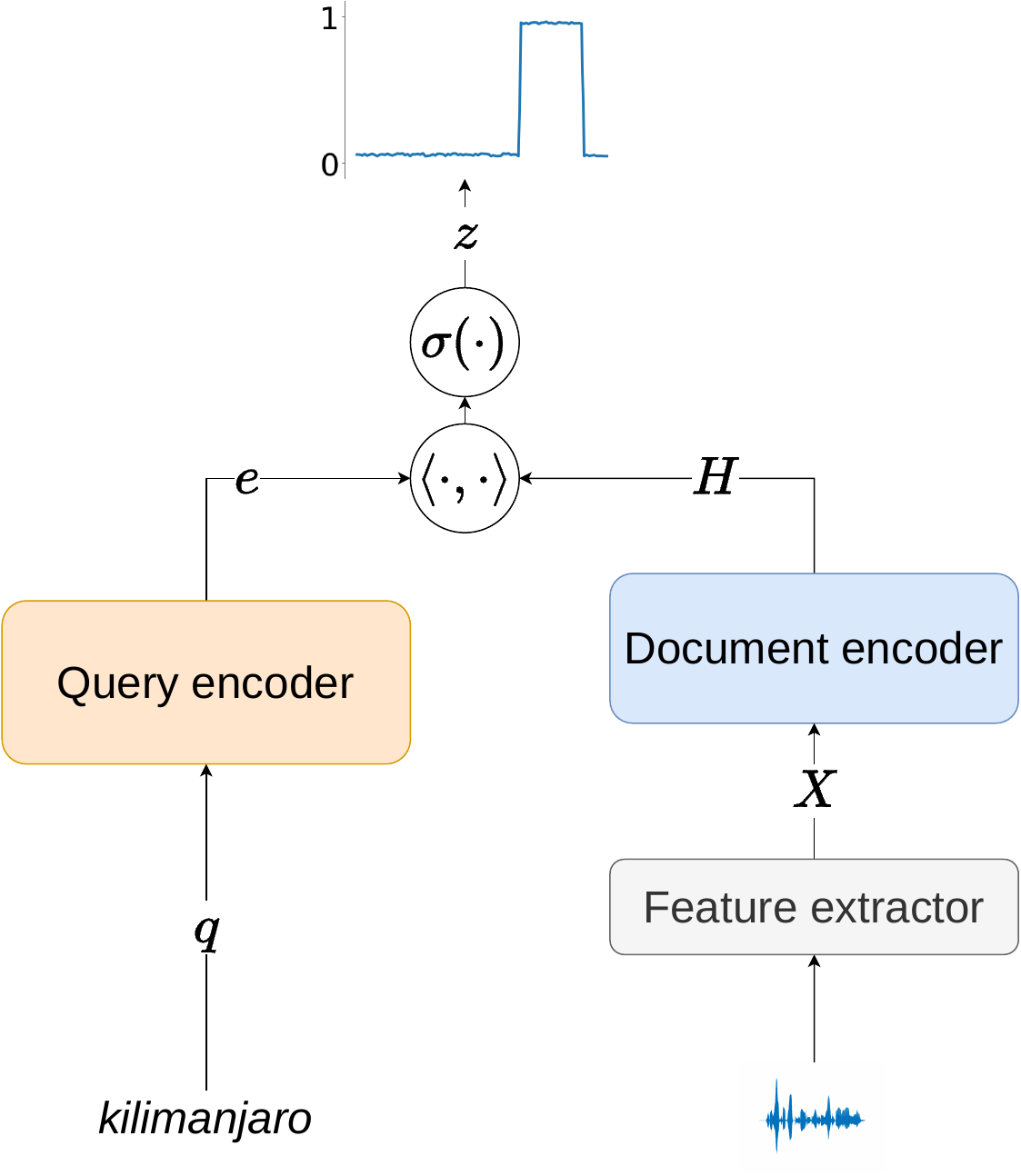}
        \caption{\baseline{} model.}
        \label{fig:baseline}
    \end{subfigure}
    \hfill
    \begin{subfigure}[b]{0.5\textwidth}
        \centering
        \includegraphics[width=\linewidth]{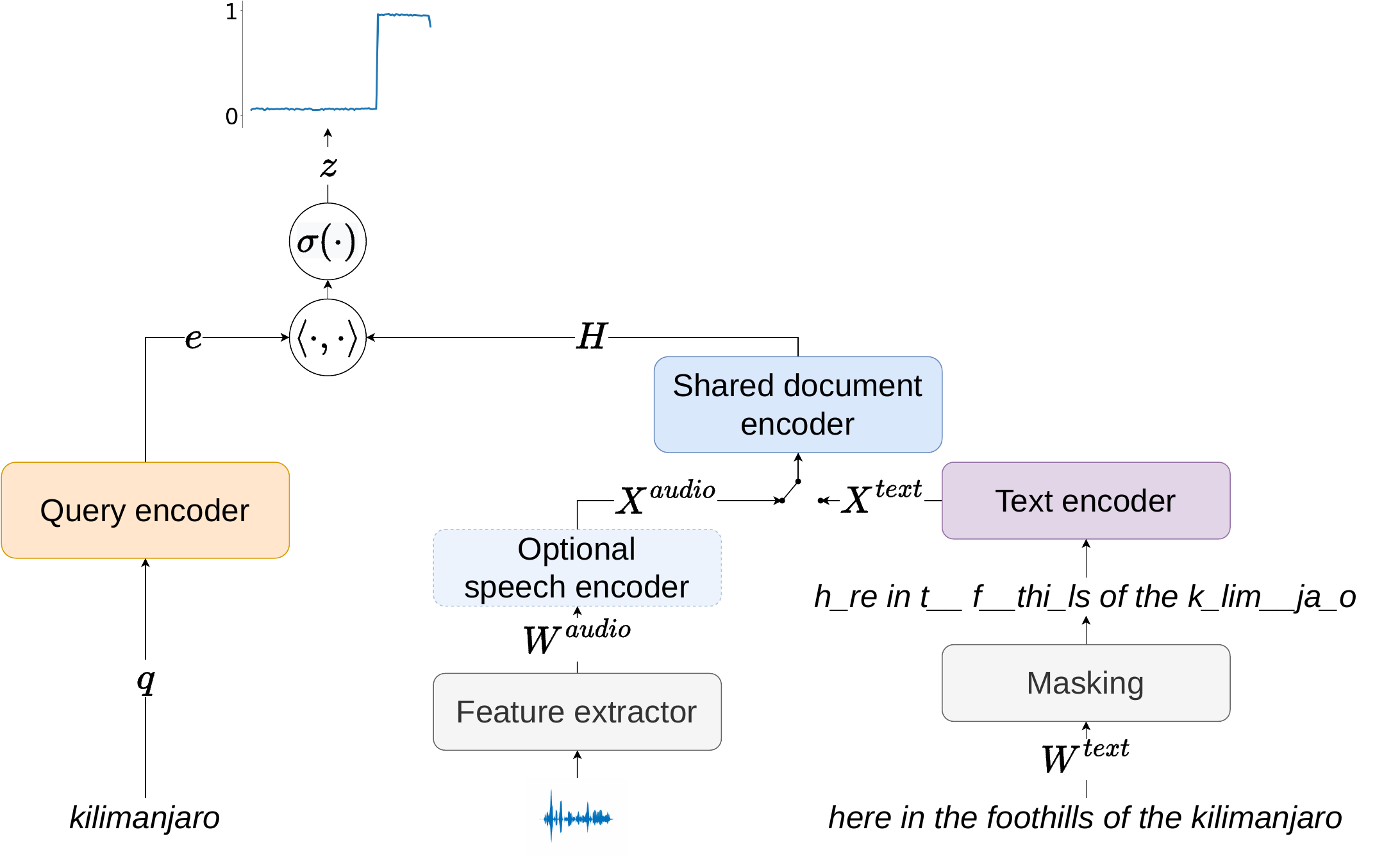}
        \caption{Proposed \proposed{} model.}
        \label{fig:proposed}
    \end{subfigure}
    \caption{Illustrations of the baseline and proposed systems.
    Both accept a written query and a spoken document and return a sequence of probabilities indicating where, if anywhere, in the document the query is spoken. The proposed system, however, also accepts documents in text form through a text encoder, thereby allowing the possibility of training with text-only data.
    }
    \label{fig:model}
\end{figure*}

In this section, we describe the joint training method that we propose for KWS.
First, in Section~\ref{sec:methods:baseline}, we recapitulate the baseline E2E KWS method from~\cite{yusuf2023end}---which we will subsequently refer to as baseline end-to-end KWS (\baseline)---as it forms the basis of our method.
Then in Section~\ref{sec:methods:proposed}, we describe the proposed joint model, which we will subsequently refer to as the joint speech and text retriever (\proposed).
In Section~\ref{section:methods:training}, we provide details on how we train the model\footnote{Code for \baseline{} and \proposed{} available at https://github.com/bolajiy/golden-retriever}.

\subsection{Baseline end-to-end keyword search (\baseline{})}
\label{sec:methods:baseline}
\baseline{}---depicted in Figure~\ref{fig:baseline}---is a model trained to predict the probabilities of a query occurring in each frame of a spoken document.
For a (possibly multi-word) query $\Matrix{q} = \left(q_1, q_2, \dots, q_K\right)$ comprising a sequence of $K$ letters and a document $\Matrix{X} = \left(\Matrix{x}_1, \Matrix{x}_2, \dots, \Matrix{x}_{N}\right)$ comprising a sequence of $N$ acoustic frames, the model is used to predict the sequence $\Matrix{y}\left(\Matrix{q}, \Matrix{X}) = (y_1, \dots, y_{N}\right)$, where each $y_n \in \left\{0, 1\right\}$ is a binary random variable indicating the existence of the query, i.e:
\begin{align}
    y_n=
    \begin{cases}
      1, & \text{if}\ \Matrix{q} \text{ is spoken in } \Matrix{X} \text{ in a time span including } n \\
      0, & \text{otherwise}.
    \end{cases}
    \label{eqn:y_speech}
\end{align}
The model comprises a document encoder and a query with parameters $\Matrix{\Delta}$ and $\Matrix{\psi}$ respectively. Given the query $\Matrix{q}$ and the document $\Matrix{X}$, the model outputs the sequence $\Matrix{z}\left(\Matrix{q}, \Matrix{X}; \Matrix{\theta}\right) = \left(z_1, \dots, z_{{N}}\right)$ of query occurrence probabilities where:
\begin{align}
    z_{{n}} \left(\Matrix{q}, \Matrix{X}; \Matrix{\theta} \right) = \sigma \left( \Matrix h_{{n}}^\top \Matrix{e}\left(\Matrix{q}; {\Matrix{\psi}}\right) \right),
    \label{eqn:combine}
\end{align}
where $\Matrix{\theta} \coloneqq \left\{ \Matrix{\Delta}, \Matrix{\psi} \right\}$;
$\Matrix{h}_{{n}}$ is the ${n}$th frame of $\Matrix{H}(\Matrix{X}; {\Matrix{\Delta}})$, a down-sampled representation of $\Matrix{X}$ computed by the document encoder;
$\Matrix{e}(\Matrix{q}; {\Matrix{\psi}})$ is the vector representation of the query computed by the query encoder, and $\sigma(\cdot)$ is the logistic sigmoid function.
Thus, $z_{{n}} \left(\Matrix{q}, \Matrix{X}; \Matrix{\theta} \right)$ is interpreted as $ P_{\Matrix{\theta}} \left(y_{{n}}=1|\Matrix{q}, \Matrix{X} \right)$.

Given a training dataset containing of a set of spoken documents $\mathcal{X}$, the model is trained by stochastic gradient descent to minimize the negative log-likelihood of the indicators:
\begin{align}
    \Matrix{\theta}^* = \argmin_{\Matrix{\theta}} \sum_{\Matrix{q} \in \mathcal{}{Q}} \sum_{\Matrix{X} \in \mathcal{X}} \sum_{{n}} -\log P_{\Matrix{\theta}}\left(y_{{n}} | \Matrix{q}, \Matrix{X} \right),
    \label{eqn:llh}
\end{align}
where the training queries are taken from $\mathcal{Q}$, the set of all unigrams, bigrams, trigrams in the transcripts of $\mathcal{X}$, and the word-level timestamps required for training are obtained by forced-alignment with an HMM-GMM-based ASR system trained on the KWS training data.
In practice, the model is trained with a modified cross-entropy objective which was introduced in~\cite{yusuf21_interspeech} (and which we will recap in Section~\ref{section:methods:training}) because it has been shown to outperform the vanilla binary cross-entropy implied by~\eqref{eqn:llh}.

\subsection{Joint speech and text retriever (\proposed{})}
\label{sec:methods:proposed}
The approach we propose for incorporating unpaired text into E2E KWS, \proposed{}---depicted in Figure~\ref{fig:proposed}---involves modifying the document encoder of \baseline{} to accept not just acoustic inputs but also textual ones.
To do so, we introduce a pair of modality encoders which transform input from their respective modalities into a shared space; a speech-only encoder which takes spoken documents as input, and a text-only encoder which takes written sentences as input.
The output of either encoder can then be fed into a shared document encoder, and combined as in Equation~\ref{eqn:combine} with the output of the (shared) query encoder to obtain probabilities of occurrence of the query in either spoken or written sentences.

For a spoken document, $\Matrix{W}^{{\mathrm{audio}}} = \left(\Matrix{w}_1, \dots, \Matrix{w}_N\right)$, we compute its representation $\Matrix{X}^{{\mathrm{audio}}} \left(\Matrix{W}^{{\mathrm{audio}}}; \Matrix{\Delta}_{{\mathrm{audio}}}\right)$ by passing it through an optional speech-only encoder with parameters $\Matrix{\Delta}_{{\mathrm{audio}}}$.
Henceforth, to reduce clutter, we drop the functional form $\Matrix{X}^{\mathrm{audio}}\left(\Matrix{W}^{{\mathrm{audio}}}; \Matrix{\Delta}_{{\mathrm{audio}}}\right)$, and simply write $\Matrix{X}^{\mathrm{audio}}$ with the understanding that the dependency is implied.
Note that, with the change of the model, we have had to make a slight change in notation:
in Section~\ref{sec:methods:baseline}, $\Matrix{X}$ denoted both the sequence of acoustic features and the document encoder input (since these are identical for \baseline{}); here, $\Matrix{W}^{\mathrm{audio}}$ denotes the sequence of acoustic features, while $\Matrix{X}^{\mathrm{audio}}$ refers to the document encoder input which computed on $\Matrix{W}^{\mathrm{audio}}$.
For most of our experiments, we do not use a speech-only encoder at all.
Rather, we use the text-only encoder to project written documents to the space of acoustic features, i.e., by default, $\Matrix{\Delta}_{{\mathrm{audio}}} = \varnothing$ and $\Matrix{X}^{{\mathrm{audio}}} = \Matrix{W}^{{\mathrm{audio}}}$.

To compute the representation for a written document, $\Matrix{W}^{{\mathrm{text}}} = \left({w}_1, \dots, {w}_N\right)$, we first mask it to obtain $\tilde{\Matrix{W}}^{{\mathrm{text}}} = \left(\tilde{w}_1, \dots, \tilde{w}_N\right)$:
\begin{align}
    \tilde{w}_n=
    \begin{cases}
      \texttt{\_}, & \text{with probability } \pi, \\
      {w}_n, & \text{with probability } 1 - \pi,
    \end{cases}
    \label{eqn:masking}
\end{align}
where $\texttt{\_}$ is a special mask symbol.
Then we incorporate a rudimentary duration model transforming the input to $\hat{\Matrix{W}}^{{\mathrm{text}}} = (\hat{w}_1, \dots, \hat{w}_N)$ where each $\hat{w}_n$ is obtained by simply repeating $\tilde{w}_n$ $\rho$ times.
For instance, the phrase $\Matrix{W}^{{\mathrm{text}}} = \texttt{thecat}$ might be converted to 
$\tilde{\Matrix{W}}^{{\mathrm{text}}} = \texttt{t\_ec\_t}$, and then,
if $\rho=2$, to $\hat{\Matrix{W}}^{{\mathrm{text}}} = \texttt{tt\_\,\_eecc\_\,\_tt}$.
This final representation is then input into the text encoder---a neural network with an embedding lookup input layer---with parameters $\Matrix{\Delta}_{{\mathrm{text}}}$ to obtain $\Matrix{X}^{{\mathrm{text}}} \left(\Matrix{W}^{{\mathrm{text}}}; \Matrix{\Delta}_{{\mathrm{text}}}\right) $.
We determine the values of $\pi$ and $\rho$ experimentally (see Section~\ref{sec:experiments:text_rep} for an analysis of their impact).

Having obtained the modality-specific representations, we can use Equation~\ref{eqn:combine} to get the occurrence probabilities  $P_{\Matrix{\theta}} \left(y_{{n}} |\Matrix{q}, \Matrix{X}^{\mathrm{audio}} \right)$ or $P_{\Matrix{\theta}} \left(y_{{n}} |\Matrix{q}, \Matrix{X}^{\mathrm{text}} \right)$ for any query $\Matrix{q}$ where the parameters $\left\{ \Matrix{\Delta}, \Matrix{\psi} \right\}$ are shared by both the speech-text retrieval and the text-text retrieval.

As stated in Section~\ref{sec:methods:baseline}, for spoken documents, $\Matrix{y}(\Matrix{q}, \Matrix{X}^{\mathrm{audio}})$ is defined by whether the query is spoken at a time span of the document.
For written documents, $\Matrix{y}\left(\Matrix{q}, \Matrix{X}^{\mathrm{text}}\right)$ is defined by whether the query occurs exactly at a given location.
Using the example from above, with document sentence $\Matrix{W}^{{\mathrm{text}}} = \texttt{thecat}$ and
$\hat{\Matrix{W}}^{{\mathrm{text}}} = \texttt{tt\_\,\_eecc\_\,\_tt}$, and a query $\Matrix{q} = \texttt{cat}$,
\begin{align}
    \Matrix{y}\left(\Matrix{q}, \Matrix{X}^{\mathrm{text}}\right) = 0 0 0 0 0 0 1 1 1 1 1 1.
    \label{eqn:y_text}
\end{align}

\subsection{Training}
\label{section:methods:training}
We train the model jointly on a paired speech-text dataset, $\mathcal{X}^{{\mathrm{audio}}}$, and an unpaired text-only one, $\mathcal{X}^{{\mathrm{text}}}$, using stochastic gradient descent.
At each training step, $k$, we sample the dataset $\mu$ uniformly from $\{ {{\mathrm{audio}}}, {{\mathrm{text}}} \}$ and minimize:
\begin{align}
    J^{\mu}_k = \sum_{l=1}^{L} \sum_{m=1}^M f \Bigl(
        & \Matrix{z} \left(\Matrix{q}^{\mu}_{k, l}, \Matrix{X}^{\mu}\left(\Matrix{W}^{\mu}_{k,l,m}; \Matrix{\Delta}_{\mu}\right); \Matrix{\theta} \right), \\ \nonumber
        & \Matrix{y} \left(\Matrix{q}^{\mu}_{k, l}, \Matrix{X}^{\mu}\left(\Matrix{W}^{\mu}_{k,l,m}; \Matrix{\Delta}_{\mu}\right) \right)
    \Bigr),
    \label{eqn:single_step}
\end{align}
where $\left\{\Matrix{q}^{\mu}_{k, 1} \dots \Matrix{q}^{\mu}_{k, L} \right\}$ is a mini-batch of $L$ queries sampled randomly from the set of unigrams, bigrams and trigrams of the dataset $\mathcal{X}^{\mu}$~\footnote{Note that, as in the baseline, we consider multiple occurrences of the same training query to be distinct elements of the set, so that the probability of sampling a particular training query is directly proportional to the number of times it occurs in the training data.};
$\left\{\Matrix{W}^{\mu}_{k,l,1}, \dots, \Matrix{W}^{\mu}_{k,l,M}\right\}$ is a set of documents sampled from the dataset such that $\Matrix{W}^{\mu}_{k,l,1}$ contains $\Matrix{q}^{\mu}_{k, l}$ while the other $M-1$ documents are sampled randomly;
$\Matrix{X}^{\mu}(\cdot)$ is the output of the corresponding modality-specific encoder;
$\Matrix{z}(\cdot)$ is the model output as described by Equation~\ref{eqn:combine};
$\Matrix{y}(\cdot)$ is the ground truth as described by Equations~\ref{eqn:y_speech}~and~\ref{eqn:y_text}; and $f(\cdot)$ is the modified binary cross-entropy function defined as:
\begin{align}
    f(z, y) = -\sum_{n=1}^{{N}}& \Bigl (\mathbbm{1}_{z_n > 1 - \phi} \cdot \left(1-y_{n}\right) \log \left(1-z_n\right)
     \nonumber \\
    & + \mathbbm{1}_{z_n < \phi} \cdot \lambda \cdot y_{n} \log z_n \Bigr),
\end{align}
where $\phi$ is a hyper-parameter controlling the tolerance of the objective to easily-classified frames and $\lambda$ controls the relative weighting of positive to negative frames.

\subsection{Post-processing for keyword search}
\label{section:methods:postprocessing}
\begin{figure}
    \centering
    \includegraphics[width=0.9\linewidth]{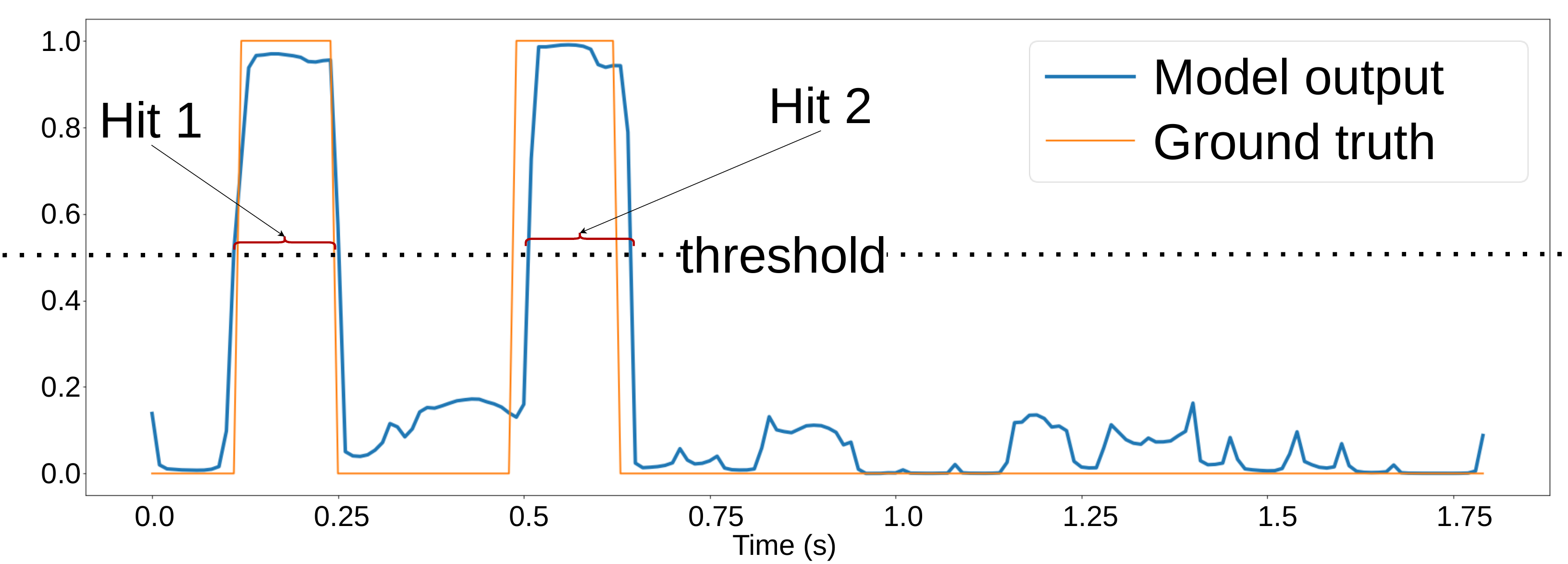}
    \caption{Post-processing of model output into KWS hypotheses.
    Contiguous regions with scores above 0.5 are selected as hits, and the confidence of each hit is the median score in corresponding region.}
    \label{fig:postprocess}
\end{figure}

After the model is trained, we no longer require the text-only document encoder $\left(\Matrix{\Delta}^{{\mathrm{text}}}\right)$, i.e., at search time, \proposed{} becomes effectively identical to \baseline{}.
We post-process the output of the query and spoken document encoders for KWS using the procedure illustrated in Figure~\ref{fig:postprocess}.
As in \baseline, for a given document, the query is detected if there exists ``islands" of consecutive frames whose sigmoid outputs, $P_{\Matrix{\theta}} \left(\Matrix{y} | \Matrix{q}, \Matrix{X}^{audio}\right)$, exceed some threshold.
We set this threshold to 0.5 in all our experiments, although we found search performance to be stable for thresholds between 0.4 and 0.7.
The first and last frame of the sequence are taken as the timestamps of the query \textit{hit} and the median probability of the sequence is taken as the confidence.
Finally, we discard hits which are shorter than 40$ms \times$query length in letters.

\section{Experiments}
\label{sec:experiments}
In this section, we conduct experiments to analyze various aspects of the proposed \proposed{} model.
First, we describe the experiment setup including datasets, input features, metrics and model configuration.
Next we present a macro comparison of the KWS performance of the proposed method to that of the \baseline{} baseline.
Then we analyze the effect of the text representation hyperparameters on search performance.
Afterwards, we conduct experiments to understand how \proposed{} achieves its improvements with analyses of the effect of the size and choice of unpaired text, the performance difference on various kinds of queries and, finally, the effect of the domain of the unpaired text on KWS performance.

\subsection{Experimental setup}
\subsubsection{Datasets}
\label{sec:experiments:datasets}
We conduct the bulk of our experiments on the IARPA Babel corpora for low resource ASR and KWS~\footnote{https://www.iarpa.gov/index.php/research-programs/babel}, from which we select Assamese~\footnote{https://catalog.ldc.upenn.edu/LDC2016S06}, Bengali~\footnote{https://catalog.ldc.upenn.edu/LDC2016S08}, Pashto~\footnote{https://catalog.ldc.upenn.edu/LDC2016S09}, Turkish~\footnote{https://catalog.ldc.upenn.edu/LDC2016S10} and Zulu~\footnote{https://catalog.ldc.upenn.edu/LDC2017S19} as the target languages for KWS training and testing.

For each language, we use the limited language pack (LLP) subset which contains about 10-hours of training data per language as the paired training data.
We use the text from the full language packs (FLP) as the unpaired text for each language.
These contain 5-6 times as many sentences as the LLP subset.

Each language has a 10-hour development (dev) set and a 5-hour evaluation (eval) set, with a few thousand queries per set.
Table~\ref{tab:text_stats} gives a summary of the text data for each language including the size of the paired text lexicon, the size of the unpaired text lexicon, and the proportion of evaluation queries which are OOV with respect to each text source.
We note that Turkish and Zulu, both agglutinative languages, have larger vocabulary sizes and higher OOV rates.

In addition to these, we use the LLP data from 19 other languages of the Babel corpus (about 190 hours in total) for multilingual pretraining of the KWS model---which was shown to significantly improve KWS performance for~\baseline{} in~\cite{yusuf2023end}--in order to measure whether and how well the proposed method can be used to improve a multilingually pretrained KWS baseline.
\begin{table}[t]
    \centering
    \caption{Statistics of the training text corpora. Vocab-L denotes the vocabulary size of the LLP (paired) training data, Vocab-F denotes the vocabulary size of the FLP (unpaired) training data, OOV-L and OOV-F refer to the proportion of evaluation queries in each language which are out-of-vocabulary of the LLP and FLP training data respectively.}
    \label{tab:text_stats}
    \begin{tabular}{lcccc}
    \toprule
    Language & Vocab-L & Vocab-F & OOV-L (\%) & OOV-F (\%) \\
    \midrule
    Assamese & 7661 & 22033 & 12.6 & 1.6\\
    Bengali & 7933 & 24339 & 13.1 & 2.1\\
    Pashto & 6186 & 17640 & 11.4 & 2.3 \\
    Turkish & 10110 & 38311 & 19.2 & 6.5\\
    Zulu &13764 & 54295 & 20.3 & 6.7\\
    \bottomrule
    \end{tabular}
\end{table}

\subsubsection{Acoustic features}
We use features from a pretrained 300 million parameter XLS-R model~\cite{babu22_interspeech} as the acoustic input to our KWS system~\footnote{https://dl.fbaipublicfiles.com/fairseq/wav2vec/xlsr2\_300m.pt}.
In a preliminary experiment on the Turkish development set, we tried using the outputs of 5th, 10th, 15th, 20th and 23rd (final) layers of the XLS-R model and found that the 15th layer worked best, and so we use it for subsequent experiments.
Note that, due to computational constraints, we only use the XLS-R model as a feature extractor rather than finetune it.

In addition to the 1024-dimensional XLS-R features, we also consider 42-dimensional multilingual bottleneck features (BNF) in Section~\ref{sec:experiments:main} as an alternative acoustic input, giving us yet another axis along which to analyze the proposed method's performance.
The BNF extractor is a TDNN-based~\cite{peddinti2015time} multilingual acoustic model which we trained in block-softmax fashion~\cite{Vesely2012} to classify clustered context-dependent triphone states on the other Babel languages' LLP data.
\begin{table*}[t]
    \caption{Term weighted value comparison between the baseline and the proposed system. Dev set results are MTWVs while eval set results are ATWVs. ``Pretrain+sp+M=8" refers to systems with multilingual pretraining, speed perturbation and increased number of negative training utterances.}
    \centering
    \begin{tabular}{l cc ccc ccc ccc ccc ccc ccc}
    \toprule
          Language &  & Pretrain+sp & \multicolumn{2}{c}{Assamese} && \multicolumn{2}{c}{Bengali} && \multicolumn{2}{c}{Pashto} && \multicolumn{2}{c}{Turkish} && \multicolumn{2}{c}{Zulu} && \multicolumn{2}{c}{Average}\\
         System &Feature&+M=8& Dev & Eval && Dev & Eval && Dev & Eval && Dev & Eval  && Dev & Eval && Dev & Eval \\
         \midrule
         \baseline{} & BNF~\cite{yusuf2023end} & \ding{55} & 17.3 & 17.9 && 18.4 & 17.0 && 13.5 & 16.3&&29.2&21.6&&21.4&22.5&&20.0&19.1 \\
         \proposed{} & BNF & \ding{55} &\textbf{22.5}&\textbf{23.5}&&\textbf{24.8}&\textbf{23.1}&&\textbf{14.9}&\textbf{18.5}&&\textbf{35.3}&\textbf{26.8}&&\textbf{25.9}&\textbf{25.7}&&\textbf{24.7}&\textbf{23.5} \\
         \midrule
         \baseline{} & XLS-R & \ding{55} &26.4&25.1&&29.8&27.4&&22.4&26.3&&41.2&34.4&&31.3&28.9&&30.2&28.4 \\
         \proposed{} & XLS-R & \ding{55} &\textbf{30.2}&\textbf{30.5}&&\textbf{34.2}&\textbf{32.7}&&\textbf{25.9}&\textbf{30.4}&&\textbf{46.6}&\textbf{39.2}&&\textbf{39.0}&\textbf{35.8}&&\textbf{35.2}&\textbf{33.7} \\
         \midrule
         \baseline{} & XLS-R & \ding{51} &34.0&34.0&&35.1&34.2&&29.9&33.4&&46.0&42.0&&39.8&36.2&&37.0&36.0 \\
         \proposed{} & XLS-R & \ding{51} &\textbf{37.9}&\textbf{37.6}&&\textbf{40.9}&\textbf{38.7}&&\textbf{31.5}&\textbf{35.2}&&\textbf{48.6}&\textbf{43.8}&&\textbf{44.4}&\textbf{42.2}&&\textbf{40.7}&\textbf{39.5} \\
         \bottomrule
    \end{tabular}
    \label{tab:main}
\end{table*}

\subsubsection{Metric}
We report the term weighted values (TWV) in all our experiments~\cite{wegmann2013tao}, which is a measure of weighted recall and precision averaged across queries.
The TWV of a set of queries $\mathcal{Q}$ at a threshold $\zeta$ is defined as:
\begin{align}
    \text{TWV}\left(\zeta, \mathcal{Q}\right) = 1 - \frac{1}{\mathcal{Q}} \sum_{q \in \mathcal{Q}} \left(P_{\text{miss}}\left(q, \zeta\right) + \beta P_{\text{FA}}\left(q, \zeta\right)\right),
    \label{eqn:twv}
\end{align}
where $P_{\text{miss}}\left(q, \zeta\right)$ is the probability of misses, $P_{\text{FA}}\left(q, \zeta\right)$ is the probability of false alarms and $\beta$ is a parameter which controls the relative importance of the two.
Following prior NIST evaluations~\cite{Fiscus2007,kws14Evalplan}, we set $\beta=999.9$.
The threshold $~\zeta$ is tuned on the dev sets.
For the dev sets, we report the maximum term weighted value (MTWV) which is the TWV at the threshold which maximizes it.
For the eval sets, we report the actual term weighted value (ATWV) which is computed by using the threshold tuned on the dev set.
Note that we report all our TWV in percentages, i.e., we always multiply the TWVs as defined in~\eqref{eqn:twv} by $100$, so $100\%$ corresponds to a perfect system with no misses and false alarms, whereas $0\%$ corresponds to a system with no outputs, and negative TWV (up to $-\beta \times 100 \%$) is possible for systems with a preponderance of false alarms.

In addition to the TWV, we report Detection Error Tradeoff (DET) curves in Section~\ref{sec:experiments:main}.
The DET curves show a plot of miss probabilities vs false alarm probabilities for a KWS system, giving a more holistic view of keyword search performance.
KWS systems with DET curves closer to the lower-left corner of the plot have better false alarm to miss tradeoffs and are thus considered better.
We use NIST's F4DE toolkit\footnote{https://github.com/usnistgov/F4DE} for computing TWVs and generating DET plots.

We adopt keyword-specific thresholding for across-query score normalization~\cite{miller2007rapid} in order to allow various queries with different score distributions to be compared with a single global threshold.

\subsubsection{Model configuration and hyper-parameters}
We base the architecture of our model on~\cite{yusuf2023end}.
The query encoder is a network with a 32-dimensional embedding layer for computing vector representations of each input grapheme, followed by 2 bidirectional gated recurrent unit (GRU) layers with 256 output units per direction per layer, and a 400-dimensional output projection layer whose outputs are summed along the sequence dimension to obtain the vectoral query representation.

The shared document encoder for \proposed{} has 6 bidirectional long short-term memory (BLSTM) layers with 512-dimensional output per direction per layer, followed by a 400-dimensional output layer.
We apply dropout of 0.4 between successive BLSTM layers, and down-sample by a factor of 2 after the fourth BLSTM layer.
By default (other than in Section~\ref{sec:experiments:layers}), we do not use any speech-only document encoder between the feature extractor and the shared encoder.
This results in document encodings with frame durations of 40ms for XLS-R features and 20ms for BNF.
The text-only document encoder comprises a 32-dimensional embedding layer, followed by a BLSTM  layer with 512-dimensional output per unit per direction, and an affine projection layer to match the input dimension of the shared encoder.

For the baseline (\baseline{}), we ensure that the configuration and number of parameters are comparable to the \proposed{} configuration above.
We use the same query encoder configuration as \proposed{} above.
We use the configuration of \proposed{}'s shared document encoder as the document encoder for \baseline{}.

For the text-document representation, we set the masking probability to $\pi=0.3$ and the duration to $\rho=2$.
We obtain these values by tuning to maximize average MTWV on Pashto, Turkish and Zulu dev sets, and apply them without tuning on Assamese and Bengali.
For the training loss function, following~\cite{yusuf2023end}, we set the positive weight to $\lambda=5$, the tolerance parameter to $\phi=0.7$ and the number of training utterances per query to $M=4$.

\subsection{Performance comparison to \baseline{}}
\label{sec:experiments:main}
\begin{figure*}[t]
    \centering
    \begin{subfigure}[b]{0.26\textwidth}
        \centering
        \includegraphics[width=\linewidth]{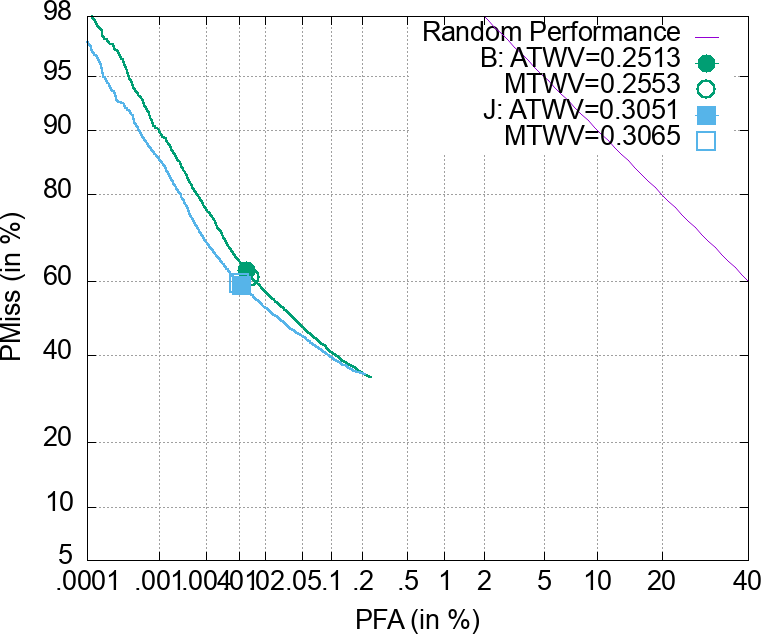}
        \caption{Assamese}
        \label{fig:assamese}
    \end{subfigure}
    \begin{subfigure}[b]{0.26\textwidth}
        \centering
        \includegraphics[width=\linewidth]{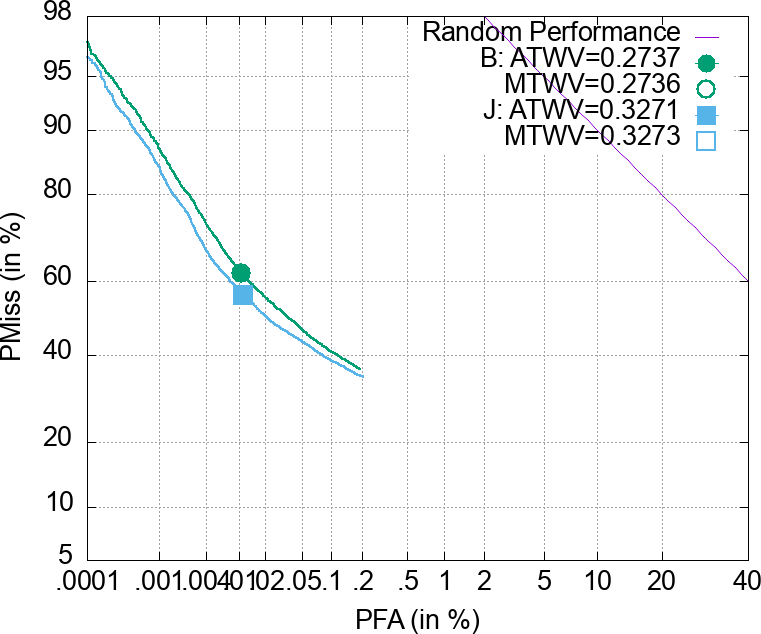}
        \caption{Bengali}
        \label{fig:bengali}
    \end{subfigure}
    \begin{subfigure}[b]{0.26\textwidth}
        \centering
        \includegraphics[width=\linewidth]{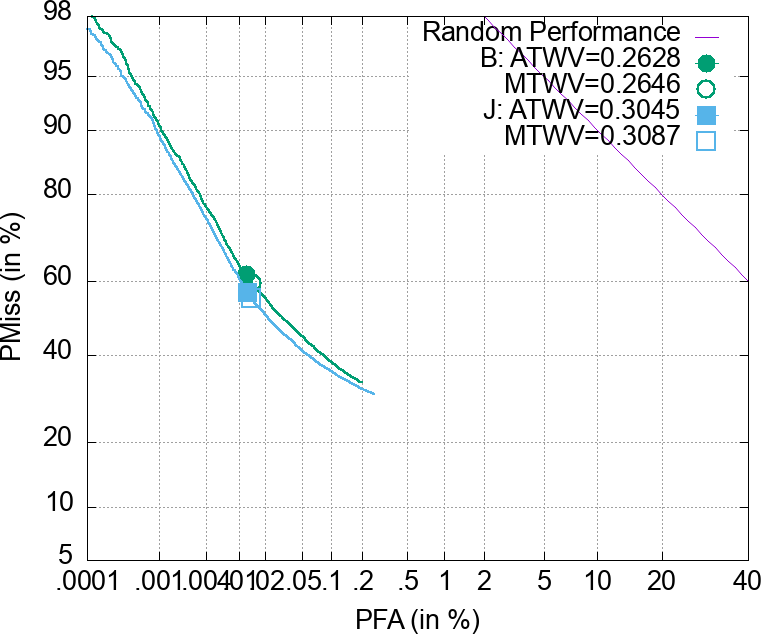}
        \caption{Pashto}
        \label{fig:pashto}
    \end{subfigure}
    \begin{subfigure}[b]{0.26\textwidth}
        \centering
        \includegraphics[width=\linewidth]{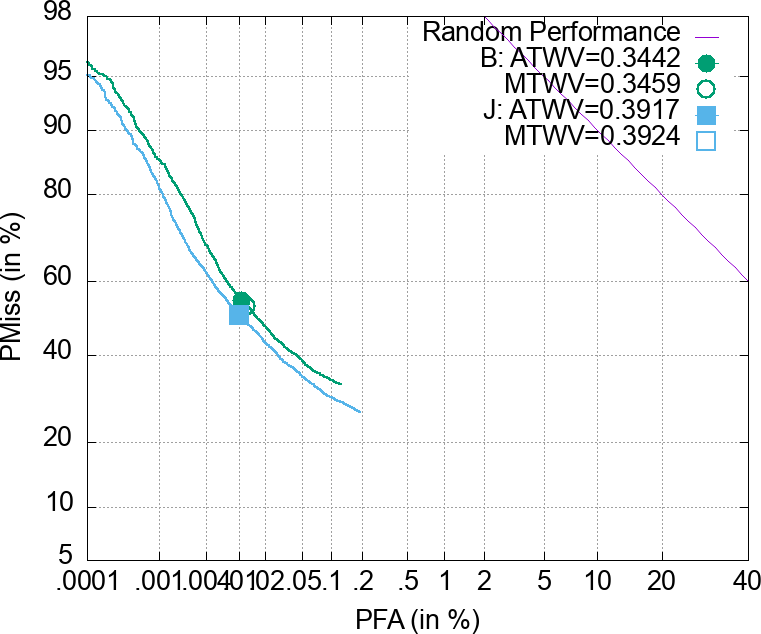}
        \caption{Turkish}
        \label{fig:turkish}
    \end{subfigure}
    \begin{subfigure}[b]{0.26\textwidth}
        \centering
        \includegraphics[width=\linewidth]{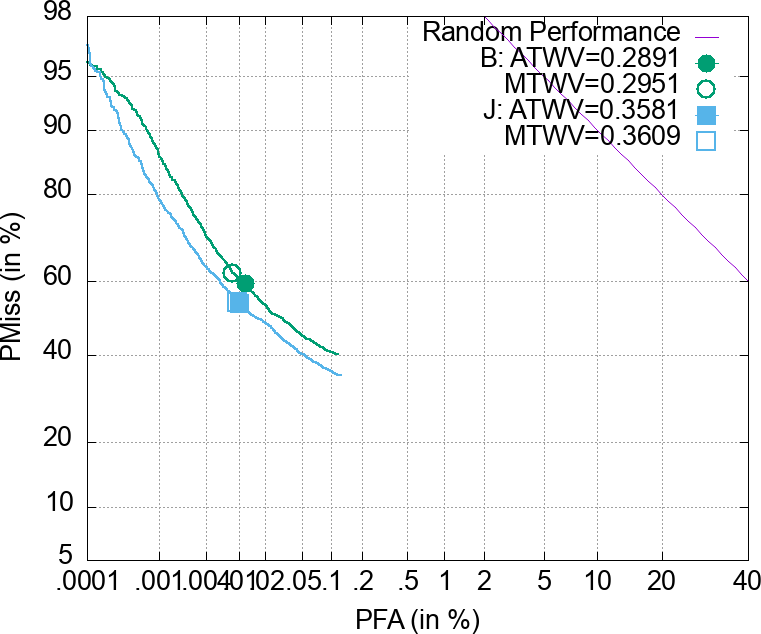}
        \caption{Zulu}
        \label{fig:zulu}
    \end{subfigure}
    \caption{DET plots showing the evolution of misses and false alarms on the evaluation sets for \baseline{} (B) and \proposed{} (J).}
    \label{fig:det_curves}
\end{figure*}
In this section, we compare the performance of \proposed{} to \baseline{} across languages and feature kinds.
Table~\ref{tab:main} shows the TWV for each of the five test languages.
For the baseline (\baseline), we note that replacing BNF as used in~\cite{yusuf2023end} with XLS-R features yields significant improvements across languages---on average, +10.2 MTWV on the dev sets and +9.3 ATWV on the eval sets.
Furthermore, by pretraining the document encoder multilingually for KWS, using speed perturbation and increasing $M$ from 4 to 8 (in Equation~\ref{eqn:single_step}), the baseline performance is increased by an additional +6.8 dev MTWV and +7.6 eval ATWV on average across languages,
showing that \baseline{} with XLS-R features can be improved with multilingual KWS pretraining despite the XLS-R features being already multilingual.
This tracks a similar finding about multilingual BNF in~\cite{yusuf2023end}.

We find that \proposed{} \textit{invariably} improves the TWV by considerable margins compared to \baseline{} in each setting (BNF, XLS-R, XLS-R + multilingual pretraining).
For BNF, the improvements across languages average +4.7 for dev set MTWV and +4.4 for eval set ATWV.
When using XLS-R features, the respective improvements increase slightly to +5 and +5.3.
When finetuning the multilingually pretrained model with XLS-R features, we get average improvements of +3.7 and +3.5 by using \proposed{} instead of \baseline{}.
Note that we use the same multilingual model---which is trained without unpaired text---to initialize the document encoders for both \baseline{} and \proposed{}.

The DET plots in Figure~\ref{fig:det_curves} provide an even more comprehensive picture of the performance difference.
In each test language, \proposed{} outperforms \baseline{} across virtually all operating points of the plots; i.e., at any given recall rate, \proposed{} incurs fewer false alarms than \baseline{}, further strengthening the significance of the superiority of \proposed{}.

\subsection{Text pre-processing}
\label{sec:experiments:text_rep}
As described in Section~\ref{sec:methods:proposed}, when computing the representation of written documents, we first mask with probability $\pi = 0.3$ and repeat each token $\rho = 2$ times.
In this section, we quantify the significance of these choices on retrieval performance.
\begin{figure}[t]
    \centering
    \resizebox{\myplotsize\linewidth}{!}{%
        \begin{tikzpicture}
            \begin{axis}[
    xlabel={$\pi$},
    ylabel={MTWV},
    xmin=0, xmax=1,
    ymin=20.0, ymax=60,
    xtick={0,0.15,0.3,0.45,0.6,0.75,0.9,1},
    xticklabels={0,0.15,0.3,0.45,0.6,0.75,0.9,1},
    legend pos=north east,
    legend style={nodes={scale=\mylegendsize, transform shape},
    legend columns=2}
]

    \addplot[color=magenta, mark=x]
    coordinates {
        (0, 31.2)
        (0.15, 32)
        (0.3, 30.2)
        (0.45, 30.4)
        (0.6, 29.3)
        (0.75, 26.7)
        (0.9, 27.4)
        (1, 25.3)
    };
    \addlegendentry{Assamese}

    \addplot[color=teal, mark=+]
    coordinates {
        (0, 33.4)
        (0.15, 35.2)
        (0.3, 34.2)
        (0.45, 35)
        (0.6, 32.5)
        (0.75, 32.1)
        (0.9, 32.4)
        (1, 29.5)
    };
    \addlegendentry{Bengali}

    \addplot[color=violet, mark=square]
    coordinates {
        (0, 26.7)
        (0.15, 27.6)
        (0.3, 25.9)
        (0.45, 25.6)
        (0.6, 23.7)
        (0.75, 22)
        (0.9, 23.7)
        (1, 24.2)
    };
    \addlegendentry{Pashto}

    \addplot[color=red, mark=triangle]
    coordinates {
        (0, 40.7)
        (0.15, 44.1)
        (0.3, 46.6)
        (0.45, 42.6)
        (0.6, 44.2)
        (0.75, 41.6)
        (0.9, 38.6)
        (1, 37.4)
    };
    \addlegendentry{Turkish}

    \addplot[color=blue, mark=o]
    coordinates {
        (0, 36.3)
        (0.15, 39.1)
        (0.3, 39)
        (0.45, 39.1)
        (0.6, 36)
        (0.75, 34)
        (0.9, 29.5)
        (1, 29.4)
    };
    \addlegendentry{Zulu}

     \addplot[color=black, mark=*, thick, line width=1.5]
    coordinates {
        (0, 33.7)
        (0.15, 35.6)
        (0.3, 35.2)
        (0.45, 34.5)
        (0.6, 33.1)
        (0.75, 31.3)
        (0.9, 30.1)
        (1, 29.2)
    };
    \addlegendentry{Average}

    \addplot[color=black, dashed]
    coordinates {
        (0, 30.2) 
        (6, 30.2)
    };
    \addlegendentry{\baseline{} average}
\end{axis}
        \end{tikzpicture}
    }
    \caption{MTWV on the development sets as the masking rate of text documents is varied.}
    \label{fig:mask}
\end{figure}
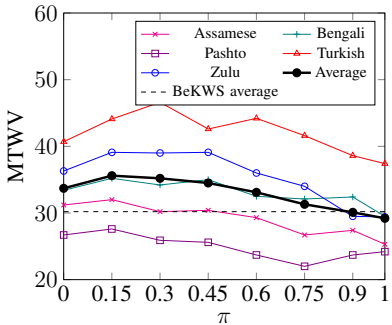

Figure~\ref{fig:mask} shows the MTWV as $\pi$ is varied with $\rho$ fixed to 2.
We find that even without masking (at $\pi=0$), \proposed{} already outperforms \baseline{} by +3.5 MTWV.
This runs counter to our original intuition that without masking, retrieval from written sentences would be too trivial to aid learning.
Nevertheless, setting $\pi$ to $0.15$ further improves MTWV by an average of +1.9.
Increasing $\pi$ further starts to worsen performance.
We note that although MTWV varies with the masking rate, only at extreme values ($\pi >0.9$) does it get worse than the baseline, indicating that the joint training is robust across a large range of $\pi$.
We surmise that having the text input is crucial, and the masking acts as extra regularization in the vein of dropout.

\begin{figure}[t]
    \centering
    \resizebox{\myplotsize\linewidth}{!}{%
        \begin{tikzpicture}
            \input{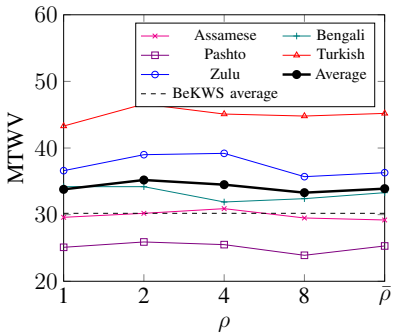}
        \end{tikzpicture}
    }
    \caption{MTWV on the development sets as the duration of each letter in text documents is varied.}
    \label{fig:trep}
\end{figure}

Figure~\ref{fig:trep} shows the performance of \proposed{} as $\rho$ is varied with $\pi$ fixed to 0.3.
\proposed{} outperforms the baseline across all the settings of $\rho$ we tried.
Although the average MTWV at $\rho=2$ is better than the MTWV at $\rho = 1$, by 1.4, the latter may still be preferred as the computational cost of the text document pipeline increases linearly with $\rho$.
Finally, we consider a more involved duration model (denoted $\bar{\rho}$ in the figure), where we set $\rho$ for each letter to be its average duration---estimated by forced-alignment with graphemic HMMs trained for each language.
We find that this added complexity yields no TWV improvements. In fact, it generally degrades performance compared to fixed duration with $1\le \rho \le 4$.

Overall, we note that although both parameters can change the performance of the system, the variance is low enough that \proposed{} still outperforms \baseline{} over large ranges of either parameter.

\subsection{\New{Number of negative utterances per training step}}
\label{sec:experiments:negatives}
\begin{figure}[t]
    \centering
    \resizebox{\myplotsize\linewidth}{!}{%
        \begin{tikzpicture}
            \input{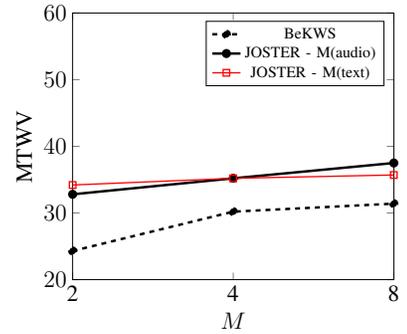}
        \end{tikzpicture}
    }
    \caption{Average MTWV on the development sets as the number of negative utterances per training step is varied.}
    \label{fig:negatives}
\end{figure}
In this section, we measure the impact of the number of negative examples in each training step.
Instead of fixing $M=4$ for both paired and unpaired batches, we vary them in turn:
\begin{itemize}
    \item $M(\mathrm{audio})$: We set $M=4$ for unpaired batches and vary it between 2, 4 and 8 for paired batches.
    \item $M(\mathrm{text})$: We set $M=4$ for paired batches and vary it between 2, 4 and 8 for unpaired batches.
\end{itemize}

Figure~\ref{fig:negatives} shows the impact as of these variations.
In both cases, we find that increasing $M$ increases the MTWV, with $M(\mathrm{audio})$ having higher impact compared to $M(\mathrm{text})$.
This however comes at the cost of increased compute and memory cost for each training step.
Note that in all our experiments, negative training utterances are sampled randomly.
We hypothesize that better sampling of negatives could result in better training efficiency or even better search accuracy, but we leave investigation of any such sampling strategies to future work.

\subsection{Number of shared layers}
\label{sec:experiments:layers}
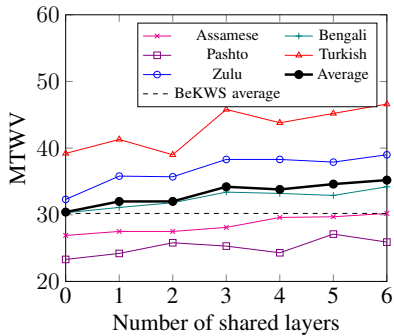
\begin{figure}[t]
    \centering
    \resizebox{\myplotsize\linewidth}{!}{%
        \begin{tikzpicture}
            \begin{axis}[
    xlabel={Number of shared layers},
    ylabel={MTWV},
    xmin=0, xmax=6,
    ymin=20.0, ymax=60,
    xtick={0,1,2,3,4,5,6},
    xticklabels={0,1,2,3,4,5,6},
    legend pos=north east,
    legend style={nodes={scale=\mylegendsize, transform shape},
    legend columns=2}
]

    \addplot[color=magenta, mark=x]
    coordinates {
        (0, 26.9)
        (1, 27.5)
        (2, 27.5)
        (3, 28.1)
        (4, 29.6)
        (5, 29.7)
        (6, 30.2)
    };
    \addlegendentry{Assamese}

    \addplot[color=teal, mark=+]
    coordinates {
        (0, 30.3)
        (1, 31.1)
        (2, 31.8)
        (3, 33.4)
        (4, 33.2)
        (5, 32.9)
        (6, 34.2)
    };
    \addlegendentry{Bengali}

    \addplot[color=violet, mark=square]
    coordinates {
        (0, 23.3) 
        (1, 24.2)
        (2, 25.8)
        (3, 25.3)
        (4, 24.3)
        (5, 27.1)
        (6, 25.9)
    };
    \addlegendentry{Pashto}

    \addplot[color=red, mark=triangle]
    coordinates {
        (0, 39.2) 
        (1, 41.3)
        (2, 39.0)
        (3, 45.8)
        (4, 43.8)
        (5, 45.2)
        (6, 46.6)
    };
    \addlegendentry{Turkish}

    \addplot[color=blue, mark=o]
    coordinates {
        (0, 32.3) 
        (1, 35.8)
        (2, 35.7)
        (3, 38.3)
        (4, 38.3)
        (5, 37.9)
        (6, 39.0)
    };
    \addlegendentry{Zulu}

    \addplot[color=black, mark=*, thick, line width=1.5]
    coordinates {
        (0, 30.4) 
        (1, 32.0)
        (2, 32.0)
        (3, 34.2)
        (4, 33.8)
        (5, 34.6)
        (6, 35.2)
    };
    \addlegendentry{Average}

    \addplot[color=black, dashed]
    coordinates {
        (0, 30.2) 
        (6, 30.2)
    };
    \addlegendentry{\baseline{} average}
\end{axis}
        \end{tikzpicture}
    }
    \caption{MTWV on the development sets as the number of layers shared between the two training tasks is varied.}
    \label{fig:layers}
\end{figure}

So far, we have fed the speech features directly into the shared encoder, i.e., there are no trainable speech-only encoder parameters.
In this section, we reduce the number of shared layers.
As we reduce the number of shared layers, we increase the number of speech-specific layers (including transferring any dropout or down-sampling components) so that the architecture and number of parameters used for the spoken document pipeline ($|\Matrix{\Delta}| + |\Matrix{\Delta}_{{\mathrm{audio}}}|$) do not change.
For instance, when we remove two LSTM layers from the shared encoder, we use a two layer LSTM network with the same configuration as the speech-only encoder.
We keep the text-only encoder configuration constant throughout.

Figure~\ref{fig:layers} shows that the MTWV generally improves as more layers are shared.
It is particularly noteworthy that when no layers are shared ($\Matrix{\Delta} = \varnothing$) by the two modalities' document encoders (with the query encoder shared as always), the MTWV is almost identical to the baseline.
This indicates that the performance improvements result from using the unpaired text to improve the (acoustic) representations learned by the document encoder rather than simply having more text data for training the query encoder.

\subsection{Size of unpaired text}
\label{sec:experiments:text_size}
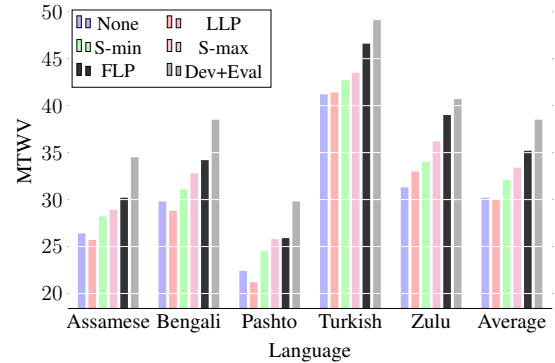
\begin{figure}[t]
    \centering
    \resizebox{0.4\textwidth}{!}{%
        \begin{tikzpicture}
            \centering
  \begin{axis}[
        ybar, axis on top,
        height=9cm, width=13cm,
        bar width=0.18cm,
        ymajorgrids,
        tick align=inside,
        major grid style={draw=white},
        axis x line*=bottom,
        xlabel={Language},
        y axis line style={opacity=0},
        tickwidth=0pt,
        enlarge x limits=true,
        legend style={
            at={(0.42, 0.95)},
            legend columns=2,
            nodes={scale=1.3, transform shape},
            /tikz/every even column/.append style={column sep=.5cm}
        },
        ylabel={MTWV},
        symbolic x coords={
           Assamese,Bengali,Pashto,Turkish,Zulu,Average},
       xtick=data,
    ]

    \addplot [draw=none, fill=blue!30] coordinates {
      (Assamese, 26.4)
      (Bengali, 29.8)
      (Pashto, 22.4)
      (Turkish, 41.2)
      (Zulu, 31.3)
      (Average, 30.2)
      };

    \addplot [draw=none, fill=red!30] coordinates {
      (Assamese, 25.7)
      (Bengali, 28.8)
      (Pashto, 21.2)
      (Turkish, 41.4)
      (Zulu, 33.0)
      (Average, 30.0)
      };

    \addplot [draw=none, fill=green!30] coordinates {
      (Assamese, 28.2)
      (Bengali, 31.1)
      (Pashto, 24.5)
      (Turkish, 42.7)
      (Zulu, 34.0)
      (Average, 32.1)
      };

    \addplot [draw=none, fill=magenta!30] coordinates {
      (Assamese, 28.9)
      (Bengali, 32.8)
      (Pashto, 25.8)
      (Turkish, 43.5)
      (Zulu, 36.2)
      (Average, 33.4)
      };

    \addplot [draw=none, fill=black!80] coordinates {
      (Assamese, 30.2)
      (Bengali, 34.2)
      (Pashto, 25.9)
      (Turkish, 46.6)
      (Zulu, 39.0)
      (Average, 35.2)
      };

      \addplot [draw=none, fill=black!30] coordinates {
      (Assamese, 34.5)
      (Bengali, 38.5)
      (Pashto, 29.8)
      (Turkish, 49.1)
      (Zulu, 40.7)
      (Average, 38.5)
      };

    \legend{None,LLP,S-min,S-max,FLP,Dev+Eval}
  \end{axis}
        \end{tikzpicture}
    }
    \caption{MTWV on the development sets as the size and composition of unpaired text is varied. None refers to the baseline with no unpaired text, LLP refers to using the LLP text for joint training, S-min denotes the worst of three randomly selected LLP-sized texts while S-max denotes the best of the three, FLP denotes using the entire FLP text for training.}
    \label{fig:text_size}
\end{figure}
In this section, we measure the impact as we change the amount of unpaired text used for the auxiliary task and report the results in Figure~\ref{fig:text_size}.
First, we compare using the FLP text as has been done so far to using the LLP text, i.e., using the transcripts of the paired data as the ``unpaired" text.
The LLP text performs significantly worse than using the FLP text and, in three of the five languages, worse even than \baseline{}.

Next, to test how much of this degradation is due to data size and how much of it results from using the same text, we create three random subsets of the FLP text (with the LLP text excluded) each with the same number of sentences as the LLP text.
We report the MTWV of the best (S-max) and worst (S-min) performing of these splits for each language.
We observe that even the best split performs much worse than the full FLP indicating the size of the augmentation text matters.
However, we also observe that even the worst random split outperforms the LLP text, indicating that textual diversity is also crucial.
Finally, we report a topline (Dev+Eval) where we use the text from the transcriptions of the Dev and Eval sets as the unpaired text for training, and find that, unsurprisingly, it outperforms using even the larger FLP text.
While it is unrealistic to assume that the transcription of the test set can be obtained beforehand, this shows that further improvements can be obtained if it can be somewhat anticipated.

\subsection{In-vocabulary and out-of-vocabulary queries}
\label{sec:experiments:query_type}
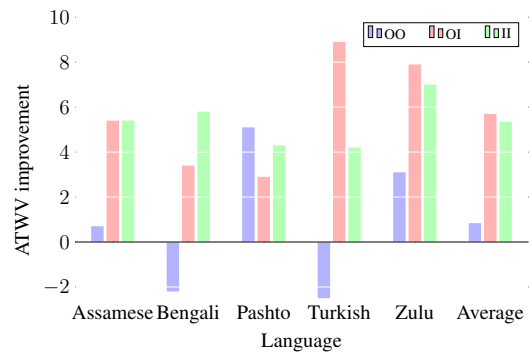
\begin{figure}[b]
    \centering
    \resizebox{0.384\textwidth}{!}{%
        \begin{tikzpicture}
            \centering
  \begin{axis}[
        ybar, axis on top,
        height=9cm, width=12.5cm,
        bar width=0.3cm,
        ymajorgrids, tick align=inside,
        major grid style={draw=white},
        axis x line*=middle,
        x tick label style={yshift={-40}},
        xlabel={Language},
        y axis line style={opacity=0},
        tickwidth=0pt,
        enlarge x limits=true,
        legend style={
            legend columns=3,
            /tikz/every even column/.append style={column sep=.5cm}
        },
        ylabel={ATWV improvement},
        symbolic x coords={
           Assamese,Bengali,Pashto,Turkish,Zulu,Average},
       xtick=data,
    ]

    \addplot [draw=none, fill=blue!30] coordinates {
      (Assamese, 0.7)
      (Bengali, -2.2)
      (Pashto, 5.1)
      (Turkish, -2.5)
      (Zulu, 3.1)
      (Average, 0.84)
      };

    \addplot [draw=none, fill=red!30] coordinates {
      (Assamese, 5.4)
      (Bengali, 3.4)
      (Pashto, 2.9)
      (Turkish, 8.9)
      (Zulu, 7.9)
      (Average, 5.7)
      };

    \addplot [draw=none, fill=green!30] coordinates {
      (Assamese, 5.4)
      (Bengali, 5.8)
      (Pashto, 4.3)
      (Turkish, 4.2)
      (Zulu, 7)
      (Average, 5.34)
      };

    \legend{OO,OI,II}
  \end{axis}
        \end{tikzpicture}
    }
    \caption{ATWV differential between the proposed system and the baseline on different subsets of the eval sets' queries. OO denotes queries which are OOV of both the LLP (paired) and FLP (unpaired) training corpora, OI denotes queries which are out of the LLP vocabulary but in the FLP vocabulary and II denotes queries which are in both vocabularies.}
    \label{fig:ivoov}
\end{figure}
In this section, we quantify how much improvement we get on various queries depending on whether or not they exist in the unpaired text.

Figure~\ref{fig:ivoov} shows the ATWV difference between \proposed{} and \baseline{} across languages for queries that are:
\begin{enumerate}
    \item OO: Out of vocabulary of both the KWS training data and the unpaired FLP text
    \item OI: Out of vocabulary with respect to the KWS training data but in the FLP text vocabulary
    \item II: In vocabulary with respect to both KWS training data and the FLP text.
\end{enumerate}
Note that for multi-word queries, out-of-vocabulary means at least one of the query words is out-of-vocabulary.

The worst average improvements over \baseline{} (+0.84 on average ATWV) are achieved for OO queries (which form a minute proportion of all queries as shown in the OOV-F column of Table~\ref{tab:text_stats}), with performance even degrading for two of the five languages.
For OI and II queries, we get consistent significant improvements (+5.7 and +5.3 average ATWV respective improvements).

Overall, we infer that \proposed{} improves the document encoder's representation of words which are in the augmentation text regardless of whether or not they actually exist in spoken form in the paired data.

\subsection{Performance in mismatched domain setting}
\label{sec:experiments:domain}
\begin{table}[t]
    \centering
    \caption{TWV on the Turkish Babel and Broadcast News datasets as the paired and unpaired training data are varied.}
    \begin{tabular}{lccccccc}
        \toprule
        &Paired&Unpaired& \multicolumn{2}{c}{BNTR} && \multicolumn{2}{c}{Babel} \\
         System & speech & text & Dev & Eval && Dev & Eval\\
         \midrule
         BeKWS & Babel & - & 55.8 & 56.1 && 41.2 & 34.4 \\ 
         JOSTER & Babel & Babel & 64.1 & 64.1 && 46.6 & 39.2\\
         JOSTER & Babel & Wikipedia & 68.5 & 67.6 && 41.2 & 37.5\\
         JOSTER & Babel & BNTR & 74.6 & 74.1 && 46.5 & 40.8\\
         \midrule
         BeKWS & BNTR & - & 84.8 & 86.1&& 24.2 & 20.5  \\
         JOSTER & BNTR & Babel &86.5 &87.8&& 29.2 & 27.4 \\
         JOSTER & BNTR & Wikipedia &87.6&88.6&&26.1&25.7 \\
         JOSTER & BNTR & BNTR & 89.0&89.8&& 28.3 & 25.6 \\
         \bottomrule
    \end{tabular}
    \label{tab:rebroadcast news}
\end{table}
So far, we have trained and tested exclusively with Babel data.
In this section, we experiment with Turkish data from various domains with various configurations of paired and unpaired data.
The objective for doing so is twofold:
\begin{enumerate}
    \item To what extent is the matching domain necessary? In other words, can we improve the TWV in one domain by using text from a different domain?
    \item Is text-only domain adaptation possible? Given paired training data in one domain and a test set in another domain, how much can we improve the test set performance using unpaired text from the target domain?
\end{enumerate}

To answer these questions, we conduct experiments on two Turkish language datasets.
In addition to the Turkish Babel dataset used in previous experiments, we use Turkish Broadcast News (BNTR)~\cite{arisoy2009turkish} for KWS training and testing.

To match the training data size of the Babel LLP corpus, we use a 10-hour subset of BNTR from the VOA channel\footnote{https://catalog.ldc.upenn.edu/LDC2012S06} for training.
\New{This training set has a vocabulary size of 16464.}
We select two 10-hour subsets from the remaining BNTR data as dev and eval sets.
Since the BNTR dataset has no official keyword lists, we randomly select 1500 queries composed of equal proportions of unigrams, bigrams and trigrams for each of the dev and eval sets \New{with OOV rates of 11.7\% and 6.5\% respectively}.
We experiment with three text corpora for unpaired training: Babel FLP text, BNTR---unpaired text from the Broadcast News dataset totalling around 180 hours
and text from Turkish Wikipedia.
The first two allow us to measure the impact of using text from the test domains, while Wikipedia stands as a control corpus.

Table~\ref{tab:rebroadcast news} shows the results of training with various configurations of paired and unpaired data on the different test sets.
First, we note that the BNTR results are generally better than the Babel ones, which is to be expected as the latter contains conversational speech from a telephone channel, while the former contains news recordings of professional newscasters.

For each test set, we get improvements by using \proposed{} regardless of the unpaired text used for joint training.
However, we get the largest improvements when we use text from the test domain to augment training.
For instance, in the cross-domain setting where we train with the paired Babel data and test on BNTR, \proposed{} with the Babel FLP text improves the dev MTWV and eval ATWV by +8.3 and +8.0 respectively compared to \baseline{}.
Augmenting with Wikipedia text results in further +4.4 and +3.5 dev and eval improvements compared to using the Babel FLP text.
Finally, using BNTR (target domain) unpaired text provides further improvements of +6.1 and +6.5 compared to Wikipedia.
This final result cuts the gap to a topline of using BNTR data for training by 65\% and 60\% on the Dev and Eval sets respectively.

In the converse cross-domain setting (training with BNTR paired data and testing on Babel), we observe a similar trend where \proposed{} using Wikipedia improves on the performance of \baseline{}, with further performance gains obtained from using BNTR text, and the best performance resulting from using Babel text.
We note, however, that the performance improvements are not as dramatic in this case---likely due to the difficulty of transferring the BNTR-trained model to the difficult acoustic conditions of the Babel data.

Finally, when training and testing within the same domain, we observe that \proposed{} generally improves the TWV compared to \baseline{} even with unpaired text from other domains.
This holds even for BNTR which already has a high baseline performance.

Overall, these results add an extra dimension to the results so far, showing that the proposed method performs well, not just across languages---as shown in previous sections---but also across domains within the same language.
Furthermore, they suggest that training \proposed{} with unpaired text from a domain most improves search performance on test sets in that domain, providing a good alternative when domain-specific data is limited.

\subsection{\New{Comparison with TTS-based text augmentation}}
\label{sec:experiments:tts}
In this section, we compare our proposed method with TTS-based unpaired text integration, where we use an off-the-shelf TTS model to synthesize speech for the unpaired text and train with the resulting data.
Here, we experiment with English language corpora due the difficulty of obtaining high-quality open-source TTS systems for other languages.
Specifically, we use the 10-hour Libri-light corpus~\cite{librilight} as the paired KWS training data, and test on the standard Librispeech test splits~\cite{panayotov2015librispeech} with around 1300 randomly generated queries for each test split.
We use the Coqui xTTS system~\cite{Eren_Coqui_TTS_2021} for speech synthesis.
We use the 100-hour Librispeech training set as the unpaired data for JOSTER and TTS.
For JOSTER, we also consider training with Wikipedia text.

Table~\ref{tab:libritts} shows the results of these experiments.
JOSTER, even with Wikipedia text, improves across all dev and test sets, and further, although the best performance is achieved when the in-domain Librispeech-100 text is used.
Similarly, using TTS for data augmentation significantly improves KWS compared to \baseline{}.

Compared to JOSTER, we note that TTS performs better on the ``clean" test sets and performs worse on the more acoustically-challenging ``other" sets.
This highlights a difference between the two approaches.
JOSTER, being text-based, is more channel-agnostic and is more influenced by linguistic similarities between the unpaired text and the target.
TTS, on the other hand, is also influenced by channel match between the output of TTS (which is typically clean speech by design) and the test audio.
Although, the impact of TTS augmentation on KWS could plausibly be improved by augmentation with artificially generated noise, reverberations or room impulse responses, an in-depth exploration of TTS-based augmentation is out of the scope of this paper. Moreover, these add extra complications which JOSTER does not have.
\begin{table}[t]
    \centering
    \caption{TWV on the Libri-Light corpus.}
    \begin{tabular}{lcccccc}
        \toprule
          & Unpaired& \multicolumn{2}{c}{Clean} && \multicolumn{2}{c}{Other} \\
         System & text & Dev & Test && Dev & Test\\
         \midrule
         BeKWS & - & 73.2 & 72.7 && 62.2 & 62.3 \\
         JOSTER & Wikipedia & 79.2 & 79.8 && 67.8 & 69.5 \\
         JOSTER  & Librispeech-100 & 82.6 & 83.0 && 71.7 & 72.7 \\
         TTS & Librispeech-100 & 85.0 & 84.9 && 67.8 & 67.8 \\
         \bottomrule
    \end{tabular}
    \label{tab:libritts}
\end{table}

\section{Conclusion}
\label{sec:conclusion}
In this paper, we propose \proposed{}, a method for integrating linguistic context into end-to-end KWS by jointly training a KWS system with an auxiliary text retrieval objective on unpaired text.
Furthermore, we conduct experiments comparing the proposed method to a baseline KWS system without the auxiliary objective, and conduct analyses to better understand how the proposed method affects the baseline KWS system.
Our experiments show the following:
\begin{itemize}
    \item The proposed method significantly improves the baseline end-to-end KWS system over several languages and feature types.
    Moreover, other approaches for improving the baseline such as multilingual pretraining and speed perturbation can also be applied on top of the proposed method to yield further improvements.
    \item Despite being trained with text, the proposed method improves document (speech) representations rather than query (text) representations of phrases in the auxiliary text.
    When such phrases are searched, the performance improves regardless of whether the phrase also occurs in the paired training data.
    On the other hand, the performance on query phrases which are not in the auxiliary text does not improve---and sometimes degrades.
    \item The proposed approach improves performance even when the auxiliary text is from a different domain than the target test set.
    However, the best performance is generally achieved when the text domain matches the test set and the proposed approach shows promise as a way to do text-only domain adaptation.
\end{itemize}

A promising avenue for future work is to extend this approach to other spoken retrieval tasks such as hotword spotting and spoken question-answering for which available paired text-to-text data dwarfs paired speech-to-text data.
Another direction is to combine it with semi-supervised training methods so as to be able leverage not just unpaired text but also unpaired speech.
\New{
Finally, like other E2E-KWS systems, ours relies on inner-product based search in vector spaces, and could therefore benefit from approximate inner-product search methods such as hashing~\cite{jansen2011efficient} or vector quantization~\cite{jegou2010product,Guo2019AcceleratingLI} which allow building fast vector indexes with sub-linear memory cost capable of handling up to trillions of documents~\cite{borgeaud2022improving} to make it competitive from a deployment standpoint
}.

\section{Acknowledgments}
\label{sec:acknowledgment}
This work was partly supported by Czech Ministry of Interior project No. VJ01010108 ``ROZKAZ".
Computing on IT4I supercomputer was supported by the Ministry of Education, Youth and Sports of the Czech Republic through e-INFRA CZ (ID:90254).
Computing on the ROYAL compute server was supported by the Turkish Directorate of Strategy and Budget under the ROYAL Project (CB SBB 2019K12-149250).

\bibliographystyle{IEEEtran}
\bibliography{my_bib}

\end{document}